\documentclass[12pt]{article}

\usepackage{graphicx}

\begin{document}
\begin{center}

{\bf Logarithmic gravity model}\\
\vspace{5mm}
 S. I. Kruglov
 \footnote{serguei.krouglov@utoronto.ca}

\vspace{5mm}
\textit{Department of Physics, University of Toronto, \\60 St. Georges St.,
Toronto, ON M5S 1A7, Canada\\
Canadian Quantum Research Center, 204-3002 32 Ave Vernon,\\
 BC V1T 2L7, Canada}
\end{center}

\begin{abstract}
The modified $F(R)$ gravity theory with the function $F(R)=-(1/\beta)\ln(1-\beta R)$ is studied.
The action at small coupling $\beta$ becomes Einstein--Hilbert action. The bound on the parameter $\beta$ from local tests is $\beta\leq 2\times 10^{-6}$ cm$^2$. We find the constant curvature solutions and it was shown that the de Sitter space is unstable but a solution with zero Ricci scalar is stable. The potential and the mass of the scalar field (scalaron) are obtained in the Einstein's frame. The
slow-roll cosmological parameters are studied and e-folds number is evaluated. The critical points of autonomous equations are analyzed. The function $m(r)$ that describes the deviation from the $\Lambda$CDM model is calculated.
\end{abstract}


\section{Introduction}

The inflation and the present time acceleration of the Universe can be explained by modification of the Einstein--Hilbert (EH) action. We study the gravity model replacing the Ricci scalar $R$ in the EH action by the particular function $F(R)$. $F(R)$-gravity models are an alternative to $\Lambda$-Cold Dark Matter ($\Lambda$CDM), and are modified gravity. It is worth noting that the introduction of the cosmological constant $\Lambda$ leads to the problem of the smallness of $\Lambda$ for the description of dark energy (DE). In addition, $\Lambda$CDM model with constant $\Lambda$ describes DE in the inflationary era but not in the current-time Universe acceleration. But $F(R)$-gravity is non-stationary model that may explain primordial and late time DE.

It was shown \cite{Appleby} that to have classical and quantum stabilities the function $F(R)$ has to satisfy the conditions $F'(R)>0$, $F''(R)>0$. There are various functions $F(R)$ to describe modified gravity. Thus, the first viable models of $F(R)$-gravity were considered in  \cite{Starobinsky,Hu,Amendola,Appleby1}.

Here, we study $F(R)$-gravity model with the function $F(R)= -(1/\beta)\ln(1-\beta R)$, where $\beta$ is the coupling. Our model reproduces the general relativity at the weak curvature limit.
It is worth mentioning that $F(R)$-gravity is the higher derivative theory where an additional degree of freedom (a scalar field) presents.
The scalar field (scalaron) may play a role of the dark matter in $F(R)$-gravity models \cite{Nojiri1,Cembranos,Inagaki}.

$F(R)$-gravity is the phenomenological model that can describe the inflation and the late-time acceleration. The successful model with $R^2$ term in the Lagrangian describing the self-consistent inflation was proposed in \cite{Starobinsky1}. Different $F(R)$-gravity models were studied in \cite{Comelli1,Comelli,Garcia,Kruglov,Kruglov1,Kruglov2,Kruglov3,Kruglov4,Kruglov6,Ketov,Deser,Gates, Wohlfarth,Wohlfarth1,Nieto,Gullu,Banados,Quiros1,Pani,Herdeiro,Fabris,Makarenko}. Reviews of $F(R)$-gravity models are given in \cite{Sotiriou,Nojiri,Capozziello}. In this paper we pay attention on the Universe inflation that can solve the problem of initial conditions (the flatness problem) necessary for the Big Bang cosmology. The large-scale homogeneity and isotropy of the Universe (the horizon problem) also can be explained by initial conditions. The scalar field which appears in the Einstein's frame drives inflation from the gravitation sector. During inflation the energy density of the Universe is dominated due to the scalar field potential. The definition of inflation is an epoch when the Universe accelerates ($\ddot{a}>0$, where $a$ is the scale factor).

The structure of the paper is as follows. In Sect. 2, we propose a model of $F(R)$-gravity with the dimension
(length)$^2$ of coupling $\beta$. The bound on coupling $\beta\leq 2\times 10^{-6} \mbox{cm}^2$ is obtained.  We find the constant curvature solution corresponding to the de Sitter space. In Sect. 3, the potential and the mass of the scalaron are found in the scalar-tensor form of the model. It was shown that the de Sitter phase is unstable but the flat space is stable. The slow-roll cosmological parameters are investigated in Sect. 4.
The function $m(r)$ which describes the deviation from the $\Lambda$CDM model is calculated.
In Sect. 5 we  study critical points of autonomous equations. Sect. 6 is a summery of results obtained.

We use the signature of the metric $\eta_{\mu\nu}$=diag(-1, 1, 1, 1) and $c$=$\hbar$=1.

\section{The Model}

The action of $F(R)$-gravity in the Jordan frame is given by
\begin{equation}
S=\int d^4x\sqrt{-g}\left[\frac{M_{Pl}^2}{2}F(R)+{\cal L}_m\right],
\label{1}
\end{equation}
were $M_{Pl}^2=1/(8\pi G)=1/\kappa^2$ is the reduced Planck mass squared and ${\cal L}_m$ is the matter Lagrangian density. The action (1) represents a scalar-tensor theory in the Jordan frame. The equivalent description can be written in the Einstein's frame with a new scalar field. Here, we consider the $F(R)$-gravity model with the function
\begin{equation}
F(R)=-\frac{1}{\beta}\ln(1-\beta R),
\label{2}
\end{equation}
where $\beta$ has the dimension of (length)$^2$ and we suppose that $\beta R<1$.
Logarithmic corrected $F(R)$ gravity, which is different from (2), was studied in \cite{Sadeghi,Odintsov,Shamir,Inagaki}.
We ignore the higher order invariants $R_{\mu\nu}R^{\mu\nu}$, $R_{\mu\nu\alpha\beta}R^{\mu\nu\alpha\beta}$ in the action and analyze $F(R)$-gravity model because EH action contains only the Ricci scalar $R$.
One can verify from Eq. (2) that $\lim_{\beta\rightarrow 0}F(R)=R$ and we have at $\beta=0$ the EH action. Thus, to recover GR at low curvature regime, we imply the smallness of parameter $\beta$. One finds the Taylor series of $F(R)$-function (2) for small $\beta R$ as follows
\begin{equation}
F(R)= R+\frac{1}{2}\beta R^2+{\cal O}(R).
\label{3}
\end{equation}
Equation (3) shows that at small $\beta R$ the model under consideration gives corrections to Starobinsky's model (the $R^2$ model) \cite{Starobinsky} which describes the self-consistent inflation \cite{Appleby}. It is worth mentioning that a small deviation
from $R^2$ model may be considered in the framework of $R^p$ model with $p\approx 2$ \cite{Motohashi}.
The laboratory experiment \cite{Kapner,Naf,Berry,Eingorn} gives the bound on the function  $F''(0)\leq 2\times 10^{-6}$ cm$^2$. Making use of Eq. (3) we obtain the restriction on the coupling $\beta$:
\begin{equation}
\beta\leq 2\times 10^{-6} \mbox{cm}^2.
\label{4}
\end{equation}
Our model satisfies observational data at the bound (4) as well as GR passes local tests. The Taylor series (3) contains all powers in Ricci curvature $R$ at $\beta R < 1$, and it is different from the Starobinsky's model.

\subsection{Constant Curvature Solutions}

Positive constant curvature de Sitter solutions to field equations (in the absence of matter)
are given by \cite{Barrow}
\begin{equation}
2F(R_0)-R_0F'(R_0)=0.
\label{5}
\end{equation}
Solutions to Eq. (5) can give a description of inflation and present time DE. Making use of Eqs. (2) and (5) we obtain
\begin{equation}
2\ln(1-\beta R_0)+\frac{\beta R_0}{1-\beta R_0}=0.
\label{6}
\end{equation}
The exact solutions to Eq. (6) are given by
\begin{equation}
\beta R_0=0,~~~\mbox{and}~~\beta R_0=1-\exp\left[W\left(-\frac{1}{2\sqrt{e}}\right)+\frac{1}{2}\right]\approx 0.72,
\label{7}
\end{equation}
where $W$ is the Lambert function ($x=W(x)\exp(W(x))$). For viability of $F(R)$-gravity models the conditions of classical and quantum stabilities $F'(R)>0$, $F''(R)> 0$ have to be satisfied \cite{Appleby}. The requirement $F'(R)>0$ leads to gravity which is attractive.
This condition is satisfied in our model because $F'(R)=1/(1-\beta R)>0$ at $\beta R)<1$.
To avoid the Dolgov--Kawasaki instability \cite{Dolgov,Amendola,Amendola1} one needs the condition $F''(R)> 0$.  This condition is also satisfied in the model under consideration as $F''(R)=\beta/(1-\beta R)^2>0$ ($\beta>0$). When $F''(R)\neq 0$ the model contains a scalaron (a scalar degree of freedom). Both conditions lead to the restriction $\beta R<1$ that also follows from Eq. (2). The positive constant curvature solutions (7) lead to classical and quantum stabilities. Thus, nontrivial solution to Eq. (6) corresponds to the Schwarzschild--de Sitter spacetime. The solution $\beta R_0\approx 0.72$ to Eq. (6) can describe primordial and present DE which is future stable if the condition $F'(R_0)/F''(R_0)>R_0$ holds, where $R_0$ is the solution to Eq. (5) \cite{Muller}. One can verify that this condition leads to the requirement $\beta R_0< 0.5$. The nontrivial solution $\beta R_0\approx 0.72$
to Eq. (6) does not satisfy the condition $F'(R_0)/F''(R_0)>R_0$. Thus, constant curvature solution corresponds to unstable de Sitter spacetime and describes the inflation. We will show that constant curvature solution matches to the maximum of the effective potential in the Einstein's frame.

\section{The Scalar-Tensor Form}

\begin{figure}[h]
\includegraphics[height=4.0in,width=4.0in]{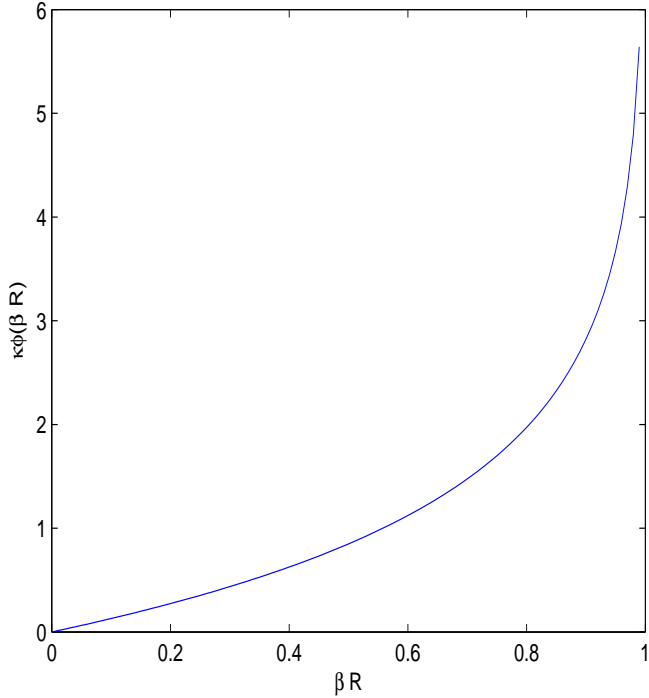}
\caption{\label{fig.1}The function $\kappa\phi$ vs $\beta R$.}
\end{figure}

\begin{figure}[h]
\includegraphics[height=4.0in,width=4.0in]{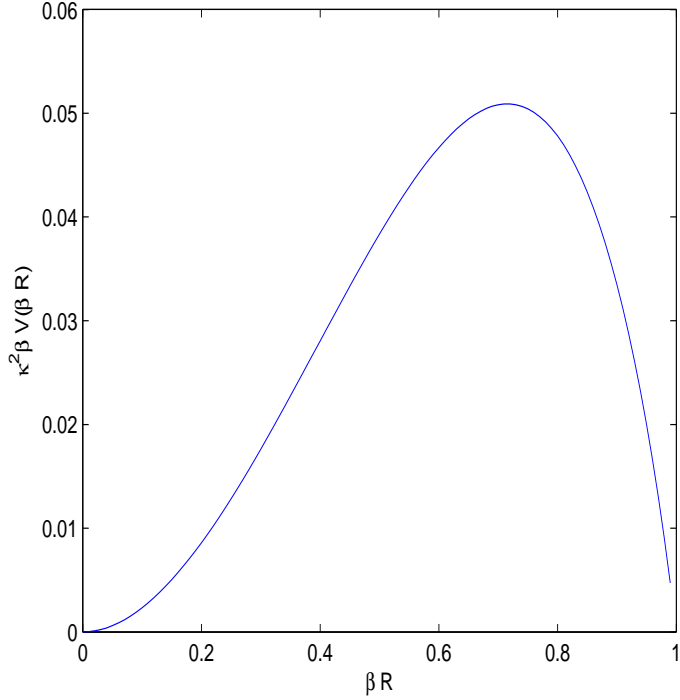}
\caption{\label{fig.2}The function $\kappa^2\beta V$ vs $\beta R$. There is the maximum at $\beta R\approx 0.7$ and the minimum at $R=0$.}
\end{figure}

The Jordan frame formulation of the modified $F(R)$-gravity can be represented in the scalar-tensor form in the Einstein's frame. Making the conformal Weyl transformation of the metric \cite{Magnano}
\begin{equation}
\widetilde{g}_{\mu\nu} =F'(R)g_{\mu\nu}=\frac{g_{\mu\nu}}{1-\beta R},
\label{8}
\end{equation}
we obtain from Eq. (1) the action of the scalar-tensor theory of gravity
\begin{equation}
S=\int d^4x\sqrt{-\widetilde{g}}\left[\frac{\widetilde{R}}{2\kappa^2}-\frac{1}{2}\widetilde{g}^{\mu\nu}
\nabla_\mu\phi\nabla_\nu\phi-V(\phi)+\widetilde{{\cal L}}_m\right].
\label{9}
\end{equation}
The Ricci scalar $\widetilde{R}$ in the Einstein's frame should be calculated by using metric (8). The scalaron field $\phi$ interacts with the  particles of the matter field in the action $\widetilde{S}_m=\int d^4x\sqrt{-\widetilde{g}}\widetilde{{\cal L}}_m$. The scalar field $\phi$ and the potential $V(\phi)$ are given by
\begin{equation}
\phi(R)=\frac{\sqrt{3}}{\sqrt{2}\kappa}\ln F'(R)=-\frac{\sqrt{3}}{\sqrt{2}\kappa}\ln (1-\beta R),
\label{10}
\end{equation}
\[
V(R)=\frac{RF'(R)-F(R)}{2\kappa^2F'^2(R)}
\]
\vspace{-7mm}
\begin{equation}
\label{11}
\end{equation}
\vspace{-7mm}
\[
=\frac{\beta R(1-\beta R)+(1-\beta R)^2\ln(1-\beta R)}{2\beta\kappa^2}.
\]
Thus, the scalar field $\phi$ is the function of the Ricci scalar $R$. It is worth noting that the energy-momentum
tensor of the matter $T^{\mu\nu}$ contributes the equation of motion for the scalaron field. Then the included effective potential is $V_{eff}=V(\phi)-V_m$, where $V_m$ depends on the trace $T^\mu_{~\mu}$ \cite{Inagaki}. The interactions
of the scalaron and matter is weak at the solar system scale because the scalaron has not been observed in the laboratory. In the presence of a matter the scalaron mass depends on the trace of the energy-momentum tensor of a matter.
In the following we study only pure gravity without a matter. The plot of the functions $\kappa\phi(\beta R)$ versus $\beta R$ is depicted in Fig \ref{fig.1}.
The function $\kappa^2\beta V(\beta R)$ versus $\beta R$ is given in Fig \ref{fig.2}.
By virtue of Eq. (11) one finds the  potential extremum
\begin{equation}
\frac{dV}{dR}=\frac{F''(R)\left[2F(R)-RF'(R)\right]}{2\kappa^2F^{'3}}=0.
\label{12}
\end{equation}
It follows from Eqs. (5) and (12) that constant curvature solutions to Eq. (6) correspond to the extremum of the potential.  Figure 2 shows that the potential (11) has the minimum at $R=0$ and the maximum at $\beta R\approx 0.72$ which are the solutions to Eq. (6). Therefore, the state corresponding to the solution $\beta R\approx 0.72$ is unstable. Making use of Eq. (11), we obtain the mass squared of a scalaron
\[
m_\phi^2=\frac{d^2V}{d\phi^2} =\frac{1}{3}\left(\frac{1}{F''(R)}+\frac{R}{F'(R)}-\frac{4F(R)}{F^{'2}(R)}\right)
\]
\begin{equation}
=\frac{1-\beta R}{3\beta}\left[1+4(1-\beta R)\ln(1-\beta R)\right].
\label{13}
\end{equation}
The plot of the function $\beta m_\phi^2$ vs $\beta R$ is depicted in Fig. 3.
\begin{figure}[h]
\includegraphics[height=4.0in,width=4.0in]{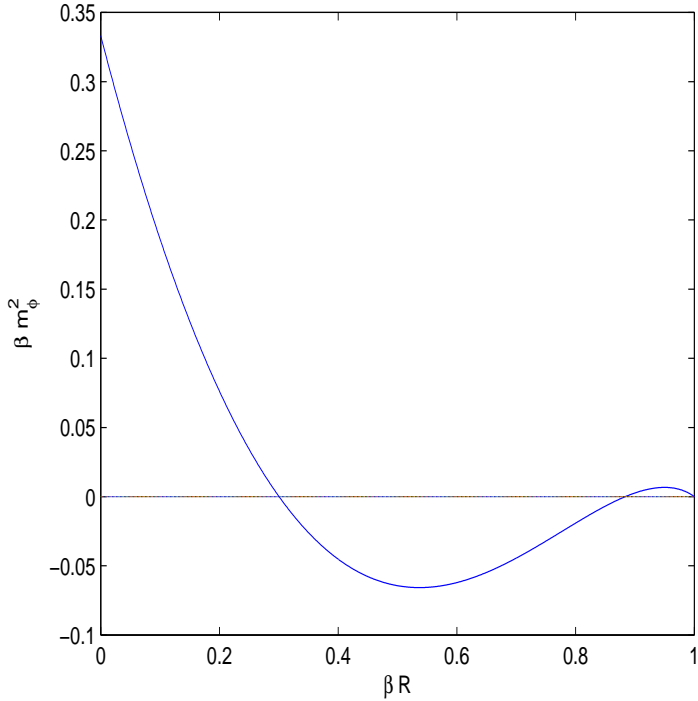}
\caption{\label{fig.3}The function $\beta m^2_\phi$ vs $\beta R$. }
\end{figure}
According to Fig. 3 the value of $m^2_\phi$ is negative ($m^2_\phi<0$) for the nontrivial constant curvature solution (7).  This again tells us that the state corresponding to the solution $\beta R\approx 0.72$ of Eq. (6) is unstable.
The stability criterion of the de Sitter solution in $F(R)$-gravity models was obtained in \cite{Muller}.  The coupling $\beta$ is small, and therefore, the squared mass $m_\varphi^2$ is big with small corrections to the Newton's law. It should be noted that matter fields give a contribution to the scalaron mass in the DE dominant era.
The solutions to equation $m_\phi=0$ are $\beta R=1$ and ($1+4(1-\beta R)\ln(1-\beta R)=0$)
\begin{equation}
\beta R_1=1-\exp\left(W\left(-\frac{1}{4}\right)\right)\simeq 0.30,~~~\beta R_2=1-\exp\left(W_{-1}\left(-\frac{1}{4}\right)\right)\simeq 0.88.
\label{14}
\end{equation}
In accordance with Fig. 3, when $\beta R_1>\beta R> 0$ or $1>\beta R>\beta R_2$ one has the stability state because $m_\phi^2>0$. When $\beta R_2> \beta R>\beta  R_1$ we have $m_\phi^2<0$ and  states are unstable.

\section{The Slow-Roll Cosmological Parameters}

Corrections to GR of F(R)-gravity model are small for $R\gg R_0$, where $R_0$ is a curvature at the present time, i.e. at high curvature regimes, if
\cite{Appleby}
\begin{equation}
\mid F(R)-R\mid \ll R,~\mid F'(R)-1\mid \ll 1,~\mid RF''(R)\mid\ll 1.
\label{15}
\end{equation}
These conditions have to be satisfied during the post-inflationary era including the radiation, matter dominated and late-time acceleration eras.
By using numerical calculations, we find that the first inequality occurs for $0.797\gg \beta R>0$. The second term in Eq. (15) leads to $0.5\gg \beta R>0$. The last inequality in Eq. (15) gives $(3-\sqrt{5})/2\gg \beta R>0$. As a result, Eq. (15) leads to
$0<\beta R\ll (3-\sqrt{5})/2\approx 0.38$. Then there is the stable Newtonian limit for all values of $R$.

The slow-roll parameters are given by \cite{Liddle}
\begin{equation}
\epsilon(\phi)=\frac{1}{2\kappa^2}\left(\frac{V'(\phi)}{V(\phi)}\right)^2,~~~~\eta(\phi)=\frac{1}{\kappa^2}\frac{V''(\phi)}{V(\phi)}.
\label{16}
\end{equation}
Making use of Eqs. (11) and (16) one finds the slow-roll parameters expressed in terms of the Ricci scalar $R$
\begin{equation}
\epsilon=\frac{1}{3}\left[\frac{RF'(R)-2F(R)}{RF'(R)-F(R)}\right]^2=
\frac{1}{3}\left[\frac{x+2(1-x)\ln(1-x)}{x+(1-x)\ln(1-x)}\right]^2,
\label{17}
\end{equation}
\[
\eta=\frac{2}{3}\left[\frac{F^{'2}(R)+F''(R)\left[RF'(R)-4F(R)\right]}{F''(R)\left[RF'(R)-F(R)\right]}\right]
\]
\begin{equation}
=\frac{2\left[1+4(1-x)\ln(1-x)\right]}{3\left[x+(1-x)\ln(1-x)\right]},
\label{18}
\end{equation}
with $x=\beta R$. During the inflation the slow-roll parameters (17) and (18) have to satisfy the inequalities $\epsilon\ll 1$ and $|\eta|<1$.
The plots of $\epsilon$ and $\eta$ versus $x$ are depicted in Fig. 4.
\begin{figure}[h]
\includegraphics[height=4.0in,width=4.0in]{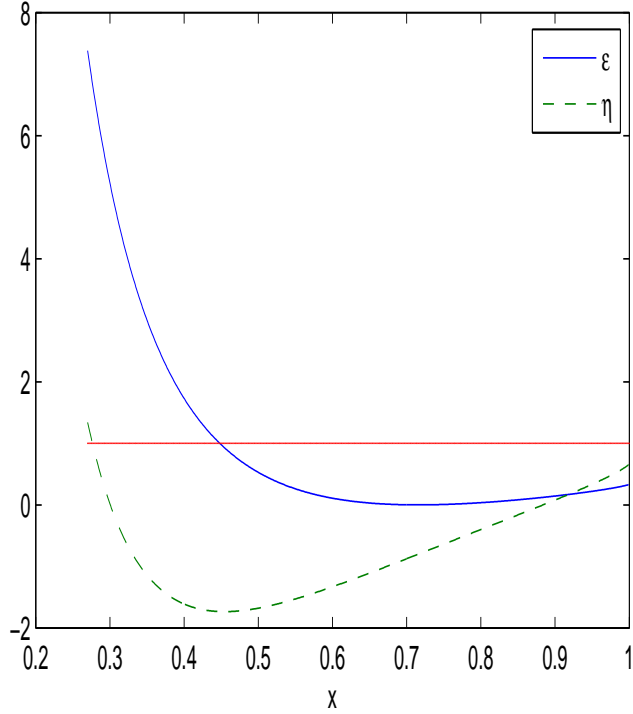}
\caption{\label{fig.4} The functions $\epsilon$ and $\eta$ vs $x=\beta R$. }
\end{figure}
The inequality $\epsilon <1$ takes place at $1>\beta R>0.48$, and $|\eta|<1$ at $0.34>\beta R>0.28$ or  $1>\beta R>0.68$. At the end of inflation
$\epsilon \simeq 1$ or $|\eta| \simeq 1$.
The age of the inflation is characterized by the $e$-folds number \cite{Liddle}
\begin{equation}
N_e\approx \kappa^2\int_{\phi_{end}}^{\phi_{start}}\frac{V(\phi)}{V'(\phi)}d\phi,
\label{19}
\end{equation}
where $\phi_{start}$ and $\phi_{end}$ correspond to the time at the start and the end of inflation. Making use od Eqs. (14) and(15) we obtain the number of $e$-foldings
\begin{equation}
N_e\approx \frac{3}{2}\int_{x_{end}}^{x_{start}}\frac{x+(1-x)\ln(1-x)}{[2(x-1)\ln(1-x)-x](1-x)}dx.
\label{20}
\end{equation}
When $\epsilon$ or $|\eta|$ are close to $1$ then value $x_{end}=\beta R_{end}$ matches to the end of inflation. It's worth mentioning that the function under the integral (20) possesses the singularity at
$2(x_0-1)\ln(1-x_0)-x_0=0$ (see Eq. (6)) with the approximate solution $x_0\simeq 0.72$. By virtue of Eq. (17) one obtains that $\epsilon=1$ at $x\simeq0.448$. We find that $|\eta|=1$ at
\[
x_1=1-\exp\left[W\left(0.2e^{0.6}\right)-0.6\right]\simeq 0.28,
\]
\[
x_2=1-\exp\left[W\left(-\frac{5}{11e^{3/11}}\right)+\frac{3}{11}\right]\simeq 0.34,
\]
\begin{equation}
x_3=1-\exp\left[W_{-1}\left(-\frac{5}{11e^{3/11}}\right)+\frac{3}{11}\right]\simeq 0.68.
\label{21}
\end{equation}
At $x_{end}=0.68$ and $x_{start}=0.7153318629591615$, one gets $N_e\approx 43$. This value is less than the amount $N_e \simeq 55-65$ which is reasonable for the inflationary era \cite{Liddle}. Because of the singularity of the function under the integral (20) one can increase the amount of inflation by increasing $x_{start}$ with the condition $x_{start}<0.72$. Our model describes the inflation but the age of the inflation is questionable.

\section{Critical Points and Stability}

To analyze the stability one introduces the dimensionless parameters \cite{Amendola}
\begin{equation}
x_1=-\frac{\dot{F}'(R)}{HF'(R)},~x_2=-\frac{F(R)}{6F'(R)H^2},~x_3=\frac{\dot{H}}{H^2}+2,
\label{22}
\end{equation}
\begin{equation}
m=\frac{RF''(R)}{F'(R)},~~~~r=-\frac{RF'(R)}{F(R)}=\frac{x_3}{x_2},
\label{23}
\end{equation}
where $H=\dot{a}/a$ is a Hubble parameter (the dot means the derivative with respect to the cosmic time) and $a(t_0) = 1$ at the present time $t_0$. Here, we assume a spatially-flat Friedmann--Lema\`{i}tre--Robertson--Walker metric, $R=6(2H^2+\dot{H})$. The function $m(r)$ describes the deviation from the $\Lambda$CDM model \cite{Amendola2}. In the absence of the radiation ($\rho_{rad}=0$), and by using variables (20) equations of motion are given by autonomous equations \cite{Amendola1}:
\begin{equation}
\frac{dx_i}{dN}=f_i(x_1,x_2,x_3),
\label{24}
\end{equation}
where $i=1,2,3$, $N=\ln a$ is the number of $e$-foldings. Functions $f_i(x_1,x_2,x_3)$ are
\[
f_1(x_1,x_2,x_3)=-1-x_3-3x_2+x_1^2-x_1x_3,
\]
\begin{equation}
f_2(x_1,x_2,x_3)=\frac{x_1x_3}{m}-x_2\left(2x_3-4-x_1\right),
\label{25}
\end{equation}
\[
f_3(x_1,x_2,x_3)=-\frac{x_1x_3}{m}-2x_3\left(x_3-2\right).
\]
The function $m(r)$ allows us to study the critical points of the equations system. By virtue of Eqs. (1) and (23), we obtain
\begin{equation}
m=\frac{x}{1-x},~~~~r=\frac{x}{(1-x)\ln(1-x)}.
\label{26}
\end{equation}
The plot of the function $m(r)$ is depicted in Fig. 5.
\begin{figure}[h]
\includegraphics[height=4.0in,width=4.0in]{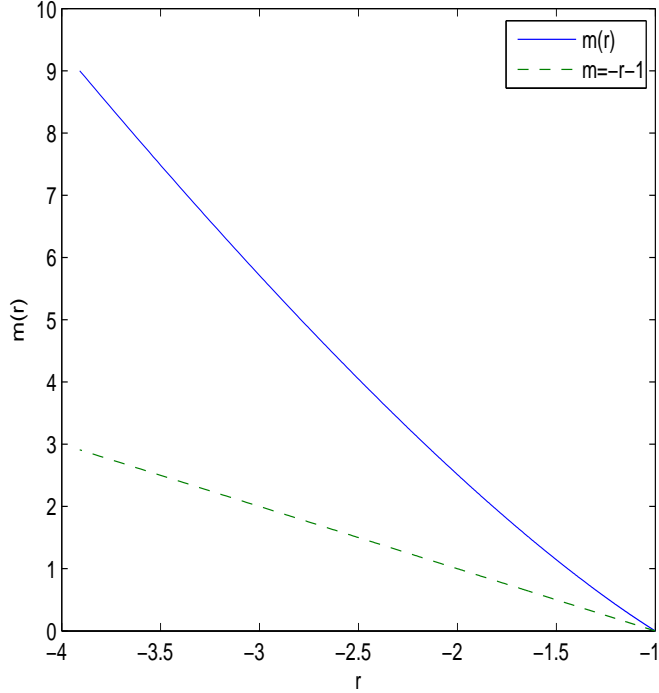}
\caption{\label{fig.5} The function $m(r)$. }
\end{figure}
The de Sitter point $P_1$ in the absence of radiation, $x_4 = 0$, is characterized by parameters $x_1=0$, $x_2=-1$, $x_3=2$. Making use of Eqs. (5), (6) and (25), we make the conclusion that this point meets the constant curvature solutions ($\dot{H}=0$). The $\Omega_{m}=1-x_1-x_2-x_3=0$ is the matter energy fraction parameter, and $w_{eff}=-1-2\dot{H}/(3H^2)=-1$ is the effective equation of state (EoS) parameter that corresponds to DE. According to Fig. 5 one has $1<m(r=-2)$ and, as a result, the constant curvature solution $x \approx 0.72$ corresponds to unstable de Sitter space-time. A viable matter dominated epoch prior to late-time acceleration exists for the critical point $P_5$ with EoS of a matter era $w_{eff}=0$, $a=a_0t^{2/3}$, $m\approx 0$, $r\approx -1$, $x_3=1/2$. The point $P_5$ belongs to the equation $m=-r-1$ with the solution $m=0$, $r=-1$ ($R=0$) (see Fig. 5). The existence of the standard matter era occurs
if the $m'(r=-1)>-1$ holds \cite{Amendola1}. With the help of Eq. (26) we obtain
\begin{equation}
\frac{dm}{dr}=\frac{\ln^2(1-x)}{x+\ln(1-x)}.
\label{27}
\end{equation}
Making use of Eq. (27) we find $\lim_{x\rightarrow 0}m'(r)=-2$. As a result, the condition $m'(r=-1)>-1$ is not satisfied and the description of the standard matter era in the model under consideration is questionable.
One needs a numerical analysis of autonomous equations to correctly describe the Universe evolution \cite{Amendola1}.

\section{Conclusion} 

We have proposed and analysed a particular $F(R)$-gravity model with the de Sitter solution that describes the Universe inflation. This model gives some corrections to the Starobinsky's $R^2$ model.
An additional degree of freedom in the Einstein's frame (the scalaron field) is responsible for the primordial inflation. The bound on the coupling $\beta$ was obtained from the local tests. It was shown that the de Sitter spacetime is unstable but the zero curvature solution is stable. The action approaches the EH action at small curvatures. Our model describes DE dynamically in the framework of $F(R)$-gravity. The potential and the mass of the scalaron were obtained in the Einstein's frame. The slow-roll parameters of the model, $\epsilon$, $\eta$, were calculated.
We studied the critical points $P_1$ and $P_5$ of autonomous equations, and the function m(r) characterizing the deviation from the $\Lambda$CDM-model has been calculated.
Note that the scalaron can be considered as a candidate for a dark matter \cite{Inagaki}. 
To describe the inflation correctly one could take into account quantum corrections. One of the ways is to add in the action some curvature invariants (for example the Gauss--Bonnet term or the Weyl tensor squared) which could mimic quantum corrections. Then, however, the  gravity will be beyond F(R)-model. Probably our model needs to be modified by adding some terms in the action to describe the primordial and late
time universe acceleration and further study.

\end{document}